\documentclass[journal]{IEEEtran}

\usepackage{graphicx}
\usepackage{amsmath, amssymb}
\usepackage{cite}
\usepackage{array}
\usepackage{multirow}
\usepackage{booktabs}
\usepackage{caption}
\usepackage{multirow}
\usepackage{tabularx}
\usepackage{float}
\usepackage{makecell}
\usepackage{float}
\usepackage{xcolor}

\usepackage{placeins}
\usepackage{algorithm}
\usepackage{algpseudocode}
\usepackage[caption=false,font=footnotesize]{subfig}

\begin{document}

\title{
UNION: A Unified AC-OPF Framework for Topology-Varying Real-Time Grid Operation
}

\author{
    Kyungnam~Park,
    Keunju~Song,
    Yeji~Lim,
    Suho~Park,
    Kibaek~Kim,
    and~Hongseok~Kim,~\IEEEmembership{Senior Member,~IEEE}%
\thanks{This work has been submitted to the IEEE for possible publication.
Copyright may be transferred without notice, after which this version may
no longer be accessible.}%
}

\maketitle

\begin{abstract}
Secure real-time grid operation requires fast AC optimal power flow (AC-OPF)
tools that stay accurate and feasible as operating conditions and topology change. Learning-based methods have advanced, but most are
trained per system or per topology, and delivering an operating point that
satisfies every operational limit remains challenging. This paper proposes UNION,
a unified graph-based AC-OPF framework for heterogeneous systems and
topology-varying operation. UNION proposes a shared graph encoder, a scalar-gated aggregation with explicit consensus correction, and a sparse-aware
differentiable implicit layer embedding the AC power-flow equations. The
remaining inequalities are handled by primal--dual training and the deterministic
restoration layer. A single model trained jointly across seven systems,
including a real-world 4,492-bus Korean transmission grid, attains a 1.23\%
mean objective gap and satisfies every operational limit on 99.56\% of test
instances. It sustains this under zero-shot \(N\!-\!1\) contingencies, i.e., line and generator
outages, and over five days of time-varying Korean topologies; it retains full
snapshot coverage at a 2.51\% gap under lightweight online fine-tuning.
UNION pre-restoration inference takes 55--58 ms per instance on the three largest systems, and
108--114 ms including restoration. These results indicate that one jointly
trained, physics-consistent model can support real-time AC-OPF across
heterogeneous systems and evolving topologies.
\end{abstract}


\section{Introduction}

\IEEEPARstart{M}{odern} power systems require fast alternating current
optimal power flow (AC-OPF) decision support for renewable variability,
rapid redispatch, and real-time security assessment under contingencies \cite{Zhang2020Powerball,Xu2012DSAWind}. AC-OPF minimizes
operating cost subject to nonlinear power-flow equations and engineering
limits, but repeatedly solving this nonconvex problem over many operating
conditions and contingencies is computationally demanding
\cite{jha2023dopfbenchmark}. Linearized approximations improve speed but
can lose AC fidelity and yield infeasible or economically misleading
decisions in stressed networks.

Learning-based surrogates amortize repeated OPF solves through offline
training, and graph neural networks (GNNs) exploit power-grid
structure. Yet many methods remain system-specific or target narrow
operating regimes, and maintaining feasibility under line and generator
outages, maintenance, and changing topologies remains difficult
\cite{Falconer2023TopologyOPF,Liu2023Topology}. Topology-flexible
and physics-guided methods improve adaptability through topology
embeddings, graph inductive biases, or equation-aware learning
\cite{yang2024pggnnopf,Zhou2023DeepOPFFT,Gao2024PGGCN}, but often
require system- or topology-specific training or do not combine
multi-system learning with topology-dependent sparse implicit computation.

We therefore propose UNION (Unified Neural Implicit Optimization Network),
a unified graph-based AC-OPF framework that jointly addresses shared
learning across heterogeneous systems, physics-consistent inference under
topology variation, and stable joint training across heterogeneous systems. Under an unseen
operating configuration, UNION predicts generator-control setpoints and
solves the nonlinear AC power-flow equations of the active topology
through a sparse differentiable implicit layer, rather than extrapolating
the full operating point. A single UNION model is trained jointly across seven systems, including the real-world 4{,}492-bus Korean transmission grid (Korea-4492), and
evaluated under normal operation, zero-shot $N\!-\!1$ line and generator
outages on four systems without contingency-specific retraining, and
five-day temporal adaptation with time-varying generator availability
and active line states.

\subsection{Literature Review}

\subsubsection{Evolution from Supervised to Physics-Informed Learning}
Learning-based AC-OPF has progressed from supervised surrogates, which
provide fast inference but depend on solver-generated labels
\cite{huang2022deepopfv}, to semi-supervised and unsupervised
formulations using pseudo-labels, data augmentation, and Lagrangian or
penalty objectives
\cite{hien2025alternative,Park2023SelfSupervised}. Although these
methods reduce supervision, penalty formulations may require careful
constraint-weight tuning and suffer from stiff trade-offs or unstable
gradients \cite{huang2024unsupervisedopf}. Physics-informed, DC-to-AC
refinement, and equation-aware methods improve physical consistency,
while DC3 and DeepLDE further incorporate feasibility completion or
equation embedding
\cite{Jiang2025PINN,Liu2026PIGCN,donti2021dc3,kim2025deeplde}.
However, scaling differentiable AC power-flow layers to graph-based
multi-system learning remains challenging because implicit
differentiation and repeated sparse solves are costly
\cite{yang2024pggnnopf,
arowolo2025generalizationgraphneuralnetworks,
piloto2024canosfastscalableneural}. Unlike system-specific
feasibility-oriented methods such as DeepOPF and QCQP-Net
\cite{pan2023deepopf,zeng2024qcqpnet}, UNION integrates
topology-aware sparse implicit differentiation into shared graph
learning, combining equation-level consistency with scalable training
across heterogeneous systems.

\subsubsection{Topology Variation and Structural Flexibility}
Operational topologies vary due to switching, maintenance, and
$N\!-\!1$ contingencies \cite{Yao2022HolomorphicCA}. Because GNNs operate on graph
representations rather than fixed-vector inputs, they naturally
support structural variation
\cite{Liu2023Topology,Donon2019Graph,Falconer2023TopologyOPF},
motivating flexible-topology predictors and topology-aware GNNs
with feasibility regularization
\cite{Zhou2023DeepOPFFT,Liu2023Topology}. However, topology
changes also modify the admittance matrix and power-flow Jacobian
sparsity, complicating differentiable sparse computation and
factorization reuse
\cite{Su2020FullParallel,Cui2021Effective}. Recent multi-system models and unified grid solvers, including LUMINA, GridSFM, and GENCO, incorporate physics through
constraint-aware training, analytical branch-flow computation, and
iterative power-balance feedback, respectively~\cite{li2026lumina,yang2026gridsfm,puech2026genco}. UNION differs
in the inference path; it predicts generator controls and uses a
sparse implicit layer to solve the nonlinear AC power-flow
equations of the active topology. Thus, after an unseen line
outage, the remaining AC state is recomputed for the modified
network. Its zero-shot $N\!-\!1$ contingency capability therefore combines
graph-based setpoint transfer with topology-conditioned physical
state completion rather than full-state GNN generalization alone.

\subsubsection{Gradient Stability in Multi-System Learning}
Joint training across heterogeneous systems is challenging because
system-wise objectives may induce conflicting update directions in
multi-objective optimization \cite{liu2025config}. In AC-OPF, this
challenge is further amplified by differences in network size,
operating conditions, and constraint scales, together with the gradient
misalignment often encountered in physics-informed optimization
\cite{wang2025gradient}. Existing approaches, including adaptive loss
balancing and dynamic task weighting, can partially mitigate such
optimization imbalances
\cite{huang2024unsupervisedopf,liu2023famo}. However, extending these
stabilization strategies to a shared graph-based AC-OPF model that incorporates an implicit physical solver remains nontrivial, particularly when
joint training spans systems with heterogeneous scales and topologies.
To address this gap, we introduce a scalar-gated aggregation interface
together with a lightweight consensus correction that acts on
system-wise gradients of the shared gate.

\subsection{Contributions}

The main contributions of UNION's integrated graph-to-solution design
are as follows.

\begin{itemize}
    \item \textbf{Scalar-gated uniform--attention aggregation:}
    We develop scalar-gated aggregation (SGA), which interpolates between
    mean pooling and learned attention pooling through a single scalar gate
    shared across systems. The gate balances uniform weighting over active
    nodes against the deviations induced by learned attention. SGA produces a fixed-dimensional, permutation-invariant graph context across
    heterogeneous network sizes and provides the scalar interface on which
    ECC operates.

    \item \textbf{Gate-level explicit consensus correction:}
    We introduce explicit consensus correction (ECC), which detects
    pairwise sign conflicts among the system-wise gradients of the shared
    SGA gate and applies a finite-difference correction to the gate
    update. This mechanism is designed to reduce cross-system
    interference without modifying the updates of the remaining model
    parameters.

    \item \textbf{Topology-conditioned sparse implicit completion:}
    We augment the predictor with a sparse-aware differentiable implicit
    layer that solves the reduced nonlinear AC power-flow equations for
    each active topology while exploiting Jacobian sparsity. The predictor
    supplies bounded generator-control setpoints, and the implicit layer
    recovers the dependent AC state. Thus, topology transfer combines
    learned setpoint transfer with topology-conditioned physical
    completion rather than full-state neural extrapolation. System-specific
    primal--dual training addresses the remaining operational inequalities.

    \item \textbf{Deterministic feasibility restoration:}
    We introduce a deterministic, training-free restoration operator that
    iteratively applies generator-side corrections and re-solves
the AC power flow to reduce residual operational-limit violations. Accepted trial
    points remain power-flow consistent, and successful restoration returns
    a point satisfying all modeled limits within the prescribed tolerance.
    Failures are retained and counted as infeasible, while the
    predictor-agnostic design enables separate pre- and post-restoration
    evaluation.

    \item \textbf{Multi-system and topology-varying validation:}
    We train one model jointly on six benchmark systems and the real-world
    Korea-4492 grid. Evaluation covers normal operation on all seven
    systems, zero-shot $N\!-\!1$ line and generator outages on four
    representative systems, and 116 hourly Korea-4492 snapshots over five
    days with changing generator availability and line states. The results
    demonstrate strong cost--feasibility performance, full temporal
    coverage under lightweight online fine-tuning, and subsecond inference
    on large-scale systems.
\end{itemize}

\section{Problem Formulation}
\label{sec:problem_formulation}

This section formulates AC-OPF for heterogeneous systems with instance-dependent active topologies, introduces the graph representation used by UNION, and states the unified learning objective.

\subsection{AC-OPF Formulation}

We consider a set $\mathcal S$ of heterogeneous power systems
and operating instances $\omega$, each associated with a system
$s\in\mathcal S$. Each instance specifies an underlying system, a load-perturbed operating condition, and an active topology. Let $\mathcal{N}_{\omega}$ and $\mathcal{E}_{\omega}\subseteq\mathcal{E}_{0}$ denote the bus set and active branch set of instance $\omega$, where $\mathcal{E}_{0}$ is the base branch set of the corresponding system. For a fixed system, $\mathcal{N}_{\omega}$ and $\mathcal{E}_{\omega}$ may vary with the node-breaker configuration \cite{Park2020Optimal}. Let $\mathcal{N}_{G,\omega}\subseteq\mathcal{N}_{\omega}$ be the
generator-bus set, and let $r\in\mathcal{N}_{G,\omega}$ denote the
reference bus. Co-located generators are aggregated per bus.

Define
\begin{equation}
\begin{aligned}
P_g &:= [P_{g,i}]_{i\in\mathcal{N}_{\omega}}, &
Q_g &:= [Q_{g,i}]_{i\in\mathcal{N}_{\omega}},\\
V &:= [V_i]_{i\in\mathcal{N}_{\omega}}, &
\theta &:= [\theta_i]_{i\in\mathcal{N}_{\omega}},
\end{aligned}
\label{eq:opf_components}
\end{equation}
and collect them into the AC-OPF decision vector
\begin{equation}
z := (P_g,Q_g,V,\theta).
\label{eq:opf_variables}
\end{equation}
For buses without generators, $P_{g,i}=Q_{g,i}=0$.

The instance parameters are denoted by
\begin{equation}
\xi_\omega :=
\bigl(
P_d,Q_d,\underline{P}_g,\overline{P}_g,\underline{Q}_g,\overline{Q}_g,
\underline{V},\overline{V},\overline{S},\mathbf{Y}_\omega,\mathbf{c}_\omega
\bigr),
\label{eq:opf_parameters}
\end{equation}
where $P_d$ and $Q_d$ are active and reactive demand vectors, $\mathbf{Y}_\omega$ is the bus-admittance matrix under the active topology, and $\mathbf{c}_\omega:=\{\mathbf{c}_i\}_{i\in\mathcal{N}_{G,\omega}}$ collects the generator cost coefficients. Underbars and overbars denote lower and upper bounds. The dependence on $\omega$ is omitted in individual symbols for readability.

Let $P_i(V,\theta)$ and $Q_i(V,\theta)$ denote the net active and reactive
power injections at bus $i$, and let $S_{ij}(V,\theta)$ denote the complex
power flow from bus $i$ to bus $j$, both given by the standard AC
network model determined by $\mathbf{Y}_{\omega}$. Transformer taps, phase
shifters, and shunt elements, when present, are absorbed into the parameters
defining $\mathbf{Y}_{\omega}$ and $S_{ij}$.

The AC-OPF problem for instance $\omega$ is written as
\begin{align}
\min_{z}\quad
& f(z;\omega):=\sum_{i\in\mathcal{N}_{G,\omega}} C_i(P_{g,i};\mathbf{c}_i)
\label{eq:opf_obj}\\
\text{s.t.}\quad
& P_{g,i}-P_{d,i}=P_i(V,\theta), && \forall i\in\mathcal{N}_{\omega},
\label{eq:pf_p}\\
& Q_{g,i}-Q_{d,i}=Q_i(V,\theta), && \forall i\in\mathcal{N}_{\omega},
\label{eq:pf_q}\\
& \underline{P}_{g,i}\le P_{g,i}\le \overline{P}_{g,i}, && \forall i\in\mathcal{N}_{G,\omega},
\label{eq:pg_bound}\\
& \underline{Q}_{g,i}\le Q_{g,i}\le \overline{Q}_{g,i}, && \forall i\in\mathcal{N}_{G,\omega},
\label{eq:qg_bound}\\
& \underline{V}_{i}\le V_i\le \overline{V}_{i}, && \forall i\in\mathcal{N}_{\omega},
\label{eq:v_bound}\\
& |S_{ij}(V,\theta)|\le \overline{S}_{ij}, && \forall \{i,j\}\in\mathcal{E}_{\omega},
\label{eq:s_bound_ij}\\
& |S_{ji}(V,\theta)|\le \overline{S}_{ij}, && \forall \{i,j\}\in\mathcal{E}_{\omega},
\label{eq:s_bound_ji}\\
& \theta_r=0.
\label{eq:slack}
\end{align}

The feasible set of instance $\omega$ is denoted by
\begin{equation}
\mathcal{Z}(\omega)
:=
\{z \mid h_{\mathrm{eq}}(z;\omega)=0,\; g_{\mathrm{ineq}}(z;\omega)\le 0\},
\label{eq:feasible_set}
\end{equation}
where $h_{\mathrm{eq}}$ collects \eqref{eq:pf_p}, \eqref{eq:pf_q}, and \eqref{eq:slack}, and $g_{\mathrm{ineq}}$ collects \eqref{eq:pg_bound}--\eqref{eq:s_bound_ji}. If the problem is feasible, let $z^\star(\omega)\in\arg\min_{z\in\mathcal{Z}(\omega)} f(z;\omega)$ denote an optimal solution.

\begin{figure*}[h]
	\centering
	\includegraphics[width=\textwidth]{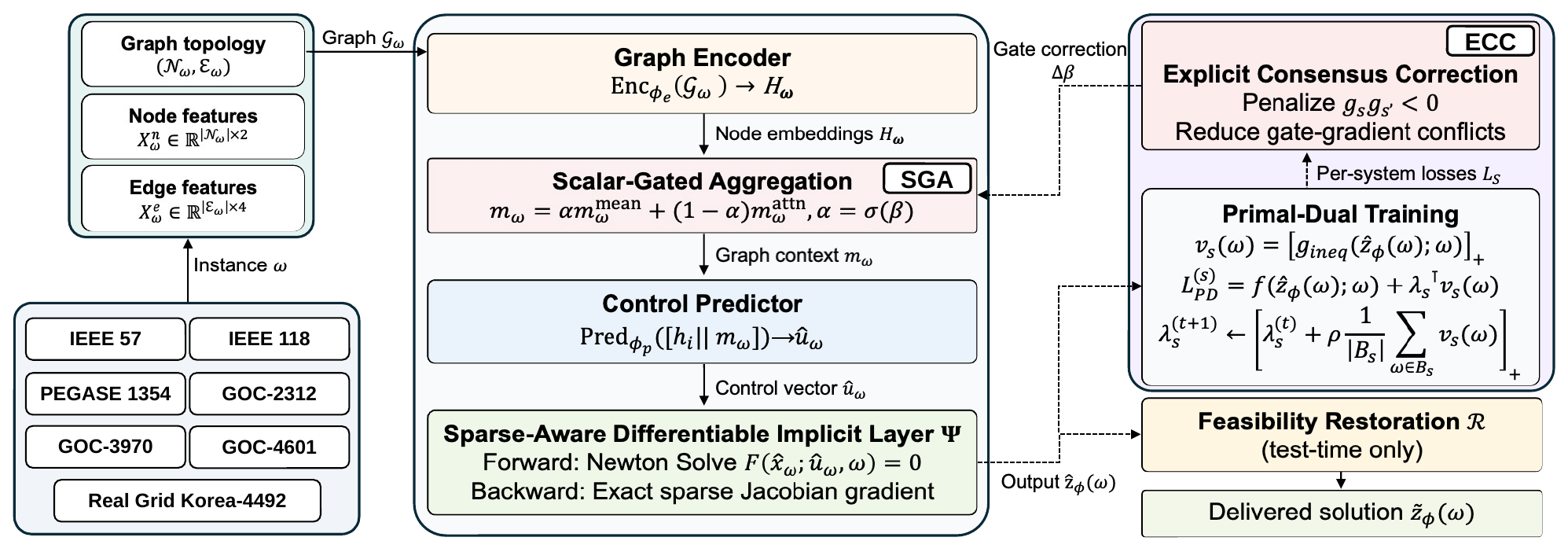}
    \caption{Overview of UNION, a unified framework for topology-varying AC-OPF.}
	\label{fig:framework}
	\vspace{-5mm}
\end{figure*}

\subsection{Graph Representation of AC-OPF Instances}
\label{subsec:graph_representation}

To support shared learning across systems with different sizes and active topologies, each AC-OPF instance is represented as an attributed graph
\begin{equation}
\mathcal{G}_{\omega} = (\mathcal{N}_{\omega}, \mathcal{E}_{\omega}, X^n_{\omega}, X^e_{\omega}),
\label{eq:graph_repr}
\end{equation}
where $X^n_{\omega}\in\mathbb{R}^{|\mathcal{N}_{\omega}|\times 2}$ contains bus active and reactive demands, and $X^e_{\omega}\in\mathbb{R}^{|\mathcal{E}_{\omega}|\times 4}$ contains branch resistance, reactance, charging capacitance, and thermal limit for each branch. Other AC-OPF data, including generator limits, voltage bounds, admittance matrices, and cost coefficients, are retained in $\xi_{\omega}$ and used in the control parameterization, implicit AC power-flow layer, training loss, and constraint evaluation.

This representation avoids fixed indexing; topology changes enter
through $\mathcal{E}_{\omega}$ and $\mathbf{Y}_{\omega}$.

\subsection{Unified Learning Objective Across Heterogeneous Systems}

Using the graph representation $\mathcal{G}_{\omega}$ and instance data $\xi_{\omega}$ defined above, our goal is to learn a shared AC-OPF predictor
\begin{equation}
\hat{z}_{\phi}(\omega)
=
\pi_{\phi}(\mathcal{G}_{\omega};\xi_{\omega})
=
(\hat{P}_g,\hat{Q}_g,\hat{V},\hat{\theta}),
\label{eq:predictor}
\end{equation}
where $\phi$ denotes the trainable parameters. Here, $\mathcal{G}_{\omega}$ is processed by the graph encoder, while
$\xi_{\omega}$ supplies the instance data used downstream. Since $\mathcal{G}_{\omega}$ is defined on the active bus and branch sets of instance $\omega$, the output dimension follows the corresponding system size and active topology.

Let $\mathcal{D}$ denote a distribution over operating instances. Formally, the unified learning problem can be written as
\begin{equation}
\begin{aligned}
\min_{\phi}\quad
& \mathbb{E}_{\omega\sim\mathcal{D}}
\bigl[f(\hat{z}_{\phi}(\omega);\omega)\bigr] \\
\text{s.t.}\quad
& \hat{z}_{\phi}(\omega)\in\mathcal{Z}(\omega),
\qquad
\forall \omega\in\operatorname{supp}(\mathcal{D}),
\end{aligned}
\label{eq:unified_learning}
\end{equation}
where $\operatorname{supp}(\mathcal{D})$ denotes the set of operating instances that can occur under $\mathcal{D}$.

This formulation captures the target behavior of UNION; a single shared predictor should produce economically competitive and AC-feasible solutions across heterogeneous systems while remaining compatible with topology-varying instances.

\section{Proposed Framework: UNION}
\label{sec:proposed_approach}

This section presents UNION as an integrated graph-to-solution system; its
architecture maps heterogeneous active graphs to power-flow-consistent
candidates, its training strategy stabilizes the shared aggregation interface
while handling the remaining operational constraints, and a restoration layer
resolves the violations that persist at inference.

\subsection{UNION Architecture}
\label{subsec:union_architecture}

The overall structure of UNION is illustrated in Fig.~\ref{fig:framework}. 
For an operating instance $\omega$ with graph representation $\mathcal{G}_{\omega}$ and instance data $\xi_{\omega}$, the shared predictor $\pi_{\phi}$ in \eqref{eq:predictor} is instantiated by the following graph-to-solution pipeline:
\begin{equation}
\mathcal{G}_{\omega}
\xrightarrow{\mathrm{Enc}_{\phi_e}}
H_{\omega}
\xrightarrow{\mathrm{SGA}_{\psi,\beta}}
m_{\omega}
\xrightarrow{\mathrm{Pred}_{\phi_p}}
\hat{u}_{\omega}
\xrightarrow{\Psi(\cdot,\omega)}
\hat{z}_{\phi}(\omega),
\label{eq:forward_pipeline}
\end{equation}
where $H_{\omega}$ is the node-embedding matrix, $m_{\omega}$ is the graph-level context vector, $\hat{u}_{\omega}$ is the predicted generator-control vector, and $\hat{z}_{\phi}(\omega)$ is the recovered AC-OPF solution \textit{candidate}. The trainable parameters are decomposed as
\begin{equation}
\phi := (\phi_e,\psi,\beta,\phi_p),
\label{eq:full_parameter_set}
\end{equation}
where $\phi_e$ parameterizes the graph encoder
$\mathrm{Enc}_{\phi_e}$, $\psi$ parameterizes the attention scoring
function in SGA, $\beta$ is its shared scalar gate logit, and
$\phi_p$ parameterizes the control predictor $\mathrm{Pred}_{\phi_p}$. The control predictor outputs generator active-power and voltage-magnitude setpoints, collected in $\hat{u}_{\omega}$. The deterministic implicit physical layer $\Psi(\cdot,\omega)$ then completes the remaining state variables by solving the reduced AC power-flow equations, yielding the final prediction $\hat{z}_{\phi}(\omega)$.

\subsubsection{Feature Embedding with EA-GNN}
\label{subsubsec:eagnn}

Using the graph representation defined in Section~\ref{subsec:graph_representation}, UNION first obtains bus-level embeddings from the active grid graph. We use the edge-aided graph neural network (EA-GNN)~\cite{song2026physicsinformedgraphlearningacceleration} as the encoder backbone. EA-GNN performs edge-aware local message passing and multi-hop graph convolution, allowing nodal demand features and active-branch attributes to be incorporated into the node embeddings. Denoting the encoder by $\mathrm{Enc}_{\phi_e}$, we obtain
\begin{equation}
H_{\omega}
=
\mathrm{Enc}_{\phi_e}(\mathcal{G}_{\omega})
\in
\mathbb{R}^{|\mathcal{N}_{\omega}|\times d_{\mathrm{emb}}},
\label{eq:encoder_output}
\end{equation}
where $d_{\mathrm{emb}}$ denotes the node-embedding dimension. The $i$th row $h_i\in\mathbb R^{d_{\mathrm{emb}}}$ embeds bus $i$.

\subsubsection{Scalar-Gated Aggregation (SGA)}
\label{subsubsec:sga}

To obtain a fixed-dimensional graph-level context from active graphs
of different sizes, we propose scalar-gated aggregation (SGA), which
combines mean and attention pooling through a single learnable scalar
gate shared across systems.

The mean-pooled representation is
\begin{equation}
m^{\mathrm{mean}}_{\omega}
=
\frac{1}{|\mathcal{N}_{\omega}|}
\sum_{i\in\mathcal{N}_{\omega}} h_i.
\label{eq:mean_pool}
\end{equation}
This normalized summary avoids the direct graph-size scaling of sum
pooling.

The attention-pooled representation uses
$\gamma_i
=
\exp(g_{\psi}(h_i))/
\sum_{j\in\mathcal{N}_{\omega}}\exp(g_{\psi}(h_j))$,
where $g_{\psi}$ is a learnable scoring function parameterized by $\psi$:
\begin{equation}
m^{\mathrm{attn}}_{\omega}
=
\sum_{i\in\mathcal{N}_{\omega}}\gamma_i h_i.
\label{eq:attn_pool}
\end{equation}
This branch emphasizes buses relevant to OPF prediction.

The graph-level context vector $m_{\omega}$ in
\eqref{eq:forward_pipeline} is then obtained as
\begin{align}
\alpha &= \sigma(\beta), \qquad \beta\in\mathbb{R},
\label{eq:scalar_gate}\\
m_{\omega}
&=
\alpha m^{\mathrm{mean}}_{\omega}
+
(1-\alpha)m^{\mathrm{attn}}_{\omega},
\label{eq:sga_context}
\end{align}
where $\sigma$ is the logistic sigmoid, $\beta$ is the shared
learnable gate logit, and $\alpha\in(0,1)$ is the mean-branch
weight. Equivalently, for $n_\omega=|\mathcal{N}_\omega|$,
\[
m_\omega
=
\sum_{i\in\mathcal{N}_\omega}w_{\omega,i}h_i,
\qquad
w_{\omega,i}
=
\frac{\alpha}{n_\omega}+(1-\alpha)\gamma_i.
\]
Hence,
$\sum_{i\in\mathcal{N}_\omega}w_{\omega,i}=1$,
$w_{\omega,i}\geq\alpha/n_\omega$, and the effective-weight
deviation from uniform pooling satisfies
$w_{\omega,i}-1/n_\omega
=(1-\alpha)(\gamma_i-1/n_\omega)$.
Thus, each active node receives the uniform component
$\alpha/n_\omega$, while $1-\alpha$ scales the
attention-induced deviation from uniform weighting.
Because both branches are fixed-dimensional and
permutation-invariant, SGA applies unchanged across graph sizes.
Moreover, the shared scalar gate logit $\beta$ provides the
one-dimensional aggregation interface on which ECC operates.

\subsubsection{Control Predictor}
\label{subsubsec:control_predictor}

Rather than directly regressing the full AC-OPF decision vector, UNION predicts a compact set of generator control setpoints, reducing the predictor output dimension while allowing the implicit layer to complete the remaining state through the AC power-flow equations.

For each generator bus $i\in\mathcal{N}_{G,\omega}$ identified from the instance data, the shared predictor $\mathrm{Pred}_{\phi_p}$ maps the local node embedding and the graph-level context to two control primitives $\tilde p_i$ and $\tilde v_i$:
\begin{equation}
[\tilde p_i,\tilde v_i]
=
\mathrm{Pred}_{\phi_p}([h_i \Vert m_{\omega}]),
\qquad i\in\mathcal{N}_{G,\omega},
\label{eq:raw_controls}
\end{equation}
where $\Vert$ denotes concatenation.

The raw outputs are mapped to instance-dependent operating ranges in $\xi_{\omega}$. For non-reference generator buses,
\begin{equation}
\hat P_{g,i}
=
\underline P_{g,i}
+
(\overline P_{g,i}-\underline P_{g,i})\sigma(\tilde p_i),
\qquad i\in\mathcal{N}_{G,\omega}\setminus\{r\},
\label{eq:pred_pg}
\end{equation}
and for all generator buses,
\begin{equation}
\hat V_i^{\mathrm{set}}
=
\underline V_i
+
(\overline V_i-\underline V_i)\sigma(\tilde v_i),
\qquad i\in\mathcal{N}_{G,\omega}.
\label{eq:pred_vset}
\end{equation}
Collecting these setpoints yields the generator control vector
\begin{equation}
\hat u_{\omega}
:=
\left(
[\hat P_{g,i}]_{i\in\mathcal{N}_{G,\omega}\setminus\{r\}},
[\hat V_i^{\mathrm{set}}]_{i\in\mathcal{N}_{G,\omega}}
\right).
\label{eq:control_vector}
\end{equation}

This parameterization enforces the active-power bounds of
non-reference generators and the voltage-setpoint bounds at
generator buses by construction. The remaining state and dependent generation quantities are
recovered by the implicit physical layer, after which branch
flows are evaluated from the recovered voltage state.

\subsubsection{Sparse-Aware Differentiable Implicit Layer}
\label{subsubsec:implicit_layer}

The generator control vector $\hat{u}_{\omega}$ does not by itself
determine all AC state variables. Following equation-embedding approaches
such as DeepLDE~\cite{kim2025deeplde}, UNION uses a differentiable implicit layer to complete the remaining state by solving the reduced AC power-flow equations, thereby embedding nonlinear power balance in the forward computation rather than relying only on residual penalties.

Let $\mathcal{N}_{PQ,\omega}\subseteq\mathcal{N}_{\omega}$ denote the set of PQ buses, and define the reduced state
\begin{equation}
x_{\omega}:=(\theta_{-r},V_{\mathcal{N}_{PQ,\omega}}),
\label{eq:implicit_state}
\end{equation}
where $\theta_{-r}$ collects the voltage angles at all buses except the slack bus $r$.

Let $\Delta P_{\omega}^{\mathrm{red}}(x_{\omega};\hat{u}_{\omega})$ and $\Delta Q_{\omega}^{\mathrm{red}}(x_{\omega};\hat{u}_{\omega})$ denote the active-power mismatches over $\mathcal{N}_{\omega}\setminus\{r\}$ and the reactive-power mismatches over $\mathcal{N}_{PQ,\omega}$, respectively. The reduced mismatch map is defined as
\begin{equation}
F(x_{\omega};\hat{u}_{\omega},\omega)
:=
\left[
\left(\Delta P_{\omega}^{\mathrm{red}}(x_{\omega};\hat{u}_{\omega})\right)^{\top}
\;
\left(\Delta Q_{\omega}^{\mathrm{red}}(x_{\omega};\hat{u}_{\omega})\right)^{\top}
\right]^{\top}.
\label{eq:reduced_mismatch_map}
\end{equation}
The full voltage and generator quantities are assembled from the reduced state $x_{\omega}$ and the control vector $\hat{u}_{\omega}$ according to the standard PV/PQ/slack-bus partition.

We denote by $\Psi$ the implicit equality-completion operator returned by the AC power-flow solve:
\begin{equation}
\hat{x}_{\omega}
=
\Psi(\hat{u}_{\omega},\omega)
\quad
\text{such that}
\quad
F(\hat{x}_{\omega};\hat{u}_{\omega},\omega)=0.
\label{eq:implicit_solve}
\end{equation}
The full AC-OPF prediction is then recovered as
\begin{equation}
\hat{z}_{\phi}(\omega)
=
\operatorname{Rec}(\hat{x}_{\omega},\hat{u}_{\omega},\omega),
\label{eq:recovery}
\end{equation}
where $\operatorname{Rec}(\cdot)$ reconstructs the remaining state and
generation quantities, with branch flows evaluated from the recovered
voltage state.

The nonlinear system in \eqref{eq:implicit_solve} is solved by
Newton--Raphson, whose steps solve
$J_x\,\Delta x_{\omega}=-F(x_{\omega};\hat{u}_{\omega},\omega)$, where
\begin{equation}
J_x(x_{\omega};\hat{u}_{\omega},\omega)
:=
\frac{\partial F(x_{\omega};\hat{u}_{\omega},\omega)}{\partial x_{\omega}}
\label{eq:state_jacobian}
\end{equation}
is the state Jacobian. The layer thus enforces nonlinear power balance up
to the solver tolerance. Gradients through $\Psi$ are computed by implicit
differentiation using a transpose solve with $J_x^{\top}$, avoiding
unrolled Newton iterations.

A direct implementation is expensive in the multi-system setting,
because sparse solves with $J_x$ and $J_x^{\top}$ are performed
repeatedly while the Jacobian sparsity pattern varies across systems
and active topologies. UNION therefore adopts the sparse-aware
execution strategy SABLE~\cite{park2026sablegpubasedpowerflow}, under
which instances sharing a sparsity pattern reuse a sparse template
whose symbolic analysis is performed once, while only its nonzero
numerical values are updated across solves.

Extending this reuse to topology-varying mini-batches is nontrivial;
a change in the active branch set invalidates the sparse template and
the associated pattern-dependent buffers and solver workspaces.
Rebuilding these objects unconditionally would eliminate the benefit
of reuse, whereas retaining them across different patterns would be
invalid. UNION therefore constructs pattern-homogeneous mini-batches
within each system and identifies the induced Jacobian sparsity
pattern at the batch level. When the pattern changes, the template
and all pattern-dependent state are rebuilt; otherwise, they are
reused with numerical-value updates only. Sparse reuse is thereby
preserved under topology changes without retaining stale
topology-specific state.

\subsection{UNION Training Strategy}
\label{subsec:training_strategy}

Training must still control the remaining operational inequality constraints and
maintain stable joint optimization across heterogeneous systems. UNION addresses both through a system-wise primal--dual objective and ECC at
the shared scalar gate, with the model parameters and the system-specific dual
variables updated jointly in a single multi-system loop.

\subsubsection{Primal--Dual Training}
\label{subsubsec:primal_dual}

The implicit layer enforces the power-balance equations up to the solver tolerance, while the control parameterization \eqref{eq:pred_pg}--\eqref{eq:pred_vset} keeps the predicted non-reference active-power and generator-voltage setpoints within their prescribed ranges. Training therefore minimizes generation cost while controlling the remaining generator reactive-power, slack-generator active-power, PQ-bus voltage, and apparent-flow limits at both branch ends.

For each system $s\in\mathcal S$, let $q_s$ denote the dimension of a fixed system-level ordering of these inequalities, and let $g_{\mathrm{ineq},s}^{\mathrm{rem}} (\hat z_{\phi}(\omega);\omega)\in\mathbb R^{q_s}$ collect the corresponding residuals, with nonpositive values denoting satisfaction and entries associated with inactive branches or unavailable components set to zero. For an instance $\omega$ from system $s$, define the nonnegative violation vector
\begin{equation}
v_s(\omega)
:=
\left[
g_{\mathrm{ineq},s}^{\mathrm{rem}}
(\hat z_{\phi}(\omega);\omega)
\right]_{+}
\in\mathbb{R}_{+}^{q_s},
\label{eq:violation_vector}
\end{equation}
where $[\cdot]_{+}:=\max(\cdot,0)$ is applied elementwise. Thus,
satisfied and inactive constraints contribute zero, and
$v_s(\omega)=0$ if and only if all remaining inequalities for
$\omega$ are satisfied.

Following primal--dual learning approaches for constrained optimization and AC-OPF \cite{Park2023SelfSupervised,Fioretto2020Predicting,kim2025deeplde}, let $\lambda_s\in\mathbb{R}_{+}^{q_s}$ be a system-specific dual vector and define the instance loss as
\begin{equation}
L_{\mathrm{PD}}^{(s)}(\phi,\lambda_s;\omega)
:=
f(\hat z_{\phi}(\omega);\omega)
+
\lambda_s^{\top}v_s(\omega),
\label{eq:primal_dual_loss}
\end{equation}
where $f(\cdot;\omega)$ is the generation cost in \eqref{eq:opf_obj}; the two terms represent cost and dual-weighted residual violations, respectively.

For a mini-batch $B_s$ sampled from system $s$, the system-wise loss is
\begin{equation}
L_s(\phi,\lambda_s)
:=
\frac{1}{|B_s|}
\sum_{\omega\in B_s}
L_{\mathrm{PD}}^{(s)}(\phi,\lambda_s;\omega),
\label{eq:system_loss}
\end{equation}
and the primal objective averaged across systems is
\begin{equation}
L_{\mathrm{train}}
(\phi;\{\lambda_s\}_{s\in\mathcal S})
:=
\frac{1}{|\mathcal S|}
\sum_{s\in\mathcal S}
L_s(\phi,\lambda_s).
\label{eq:train_objective}
\end{equation}
The standard primal gradient is computed from
$L_{\mathrm{train}}$ with the dual variables fixed, while the
system-specific dual variables are updated by projected ascent:
\begin{equation}
\lambda_s^{(t+1)}
=
\left[
\lambda_s^{(t)}
+
\rho
\frac{1}{|B_s|}
\sum_{\omega\in B_s}
v_s(\omega)
\right]_{+},
\qquad
\rho>0,
\label{eq:dual_update}
\end{equation}
where $\rho$ is the dual step size. Each $\lambda_s$ is shared across instances of system $s$, while topology-dependent inactive entries are masked in $v_s(\omega)$.

\subsubsection{Explicit Consensus Correction (ECC)}
\label{subsubsec:ecc}

The system-wise losses in \eqref{eq:system_loss} induce system-dependent update directions for the shared scalar gate $\beta$. In heterogeneous multi-system training, these directions may conflict because different systems can favor different graph-level aggregation behavior. To reduce such conflicts at the shared aggregation interface, we propose another key feature named ECC, a gate-level correction mechanism based on system-wise gate gradients.

For each training system $s\in\mathcal{S}$, the system-wise gate gradient is defined as
\begin{equation}
g_s
:=
\frac{\partial L_s(\phi,\lambda_s)}{\partial \beta},
\label{eq:gate_gradient}
\end{equation}
where the system-specific dual vector $\lambda_s$ in
\eqref{eq:system_loss} is treated as fixed when computing the primal gate gradient. We measure pairwise sign conflicts among systems through
\begin{equation}
R_{\mathrm{cons}}(\beta)
:=
\frac{2}{|\mathcal{S}|(|\mathcal{S}|-1)}
\sum_{\substack{s,s'\in\mathcal S\\ s<s'}}
\max(0,-g_s g_{s'}).
\label{eq:consensus_penalty}
\end{equation}
Here, the summation is taken over unordered pairs of distinct systems. This term is zero when all system-wise gate gradients have mutually consistent signs and becomes positive when at least one pair of systems induces opposing gate updates.

Directly differentiating $R_{\mathrm{cons}}$ would require second-order derivatives through the implicit layer, which is prohibitively expensive in the sparse-aware implementation. We therefore approximate the gate-level correction direction by finite differences. Holding $\phi_e$, $\psi$, $\phi_p$, and the dual variables $\{\lambda_s\}_{s\in\mathcal{S}}$ fixed, define
\begin{equation}
d_s
:=
\frac{g_s(\beta+\varepsilon)-g_s(\beta)}{\varepsilon},
\qquad \varepsilon>0,
\label{eq:fd_gate_gradient}
\end{equation}
which approximates $\partial g_s/\partial \beta$. The finite-difference estimate of the consensus correction direction is then
\begin{equation}
\begin{aligned}
d_{\mathrm{cons}}
:=
-\frac{2}{|\mathcal{S}|(|\mathcal{S}|-1)}
\sum_{\substack{s,s'\in\mathcal S\\ s<s'}}
\mathbb{I}[g_s g_{s'}<0]\,
\bigl(g_{s'}\,d_s + g_s\,d_{s'}\bigr),
\end{aligned}
\label{eq:ecc_fd_direction}
\end{equation}
where $\mathbb{I}[\cdot]$ is the indicator function. The shared gate logit is updated by combining the standard primal gradient step with the ECC correction:
\begin{equation}
\beta
\leftarrow
\beta
-
\eta
\frac{\partial L_{\mathrm{train}}(\phi;\{\lambda_s\}_{s\in\mathcal S})}{\partial\beta}
-
\eta_{\beta}\mu_{\mathrm{cons}}\,d_{\mathrm{cons}},
\label{eq:ecc_update}
\end{equation}
where $\eta>0$ is the standard learning rate, $\eta_{\beta}>0$ is the gate-correction learning rate, and $\mu_{\mathrm{cons}}\ge 0$ is the consensus weight. Thus, ECC modifies only the scalar gate update, while the remaining model parameters are updated by the standard primal objective.

\subsection{Feasibility Restoration Layer}
\label{subsec:restoration}

The implicit layer solves the power-balance equations up to numerical
tolerance, while primal--dual training encourages satisfaction of the
remaining inequalities; individual predictions may therefore retain
small power-balance residuals and inequality violations. To balance
computational speed and numerical accuracy, UNION executes the
predictor and implicit layer in single precision (FP32) to exploit the
target GPU's higher FP32 throughput, and uses double precision (FP64)
only for the pre-restoration refinement and subsequent restoration
solves. Hence, instead of directly using $\hat{z}_{\phi}(\omega)$ in \eqref{eq:predictor}, we consider a restored solution as follows.
Let $\bar z_{\phi}(\omega):=z_{64}(\hat u_{\omega};\omega)$ denote the
resulting power-flow-refined prediction and
$\tilde z_{\phi}(\omega):=
\mathcal R(\bar z_{\phi}(\omega),\omega)$ the delivered point.
Here, $\mathcal R$ is a deterministic, training-free restoration
operator. The initial FP64 solve uses the FP32 implicit-layer solution
as a warm start, keeps the predicted controls fixed, and terminates at
$\|F\|_{\infty}\le 10^{-8}$~p.u., thereby removing numerical
power-balance residuals without addressing inequality violations.

For a prescribed tolerance $\tau>0$, define
\begin{align}
\mathcal Z_{\tau}(\omega)
&:=
\left\{
z \,\middle|\,
h_{\mathrm{eq}}(z;\omega)=0,\;
g_{\mathrm{ineq}}(z;\omega)\le\tau\mathbf 1
\right\},
\label{eq:relaxed_feasible_set}\\
\mathcal V_{\tau}(z;\omega)
&:=
\sum_{c\in\mathcal C_{\mathrm{ineq}}(\omega)}
\left[
g_{\mathrm{ineq},c}(z;\omega)-\tau
\right]_+ .
\label{eq:violation_mass}
\end{align}
Here, $\mathcal C_{\mathrm{ineq}}(\omega)$ denotes the index set
of scalar inequality constraints for instance $\omega$.
For any power-flow-consistent point $z$,
$\mathcal V_{\tau}(z;\omega)=0$ if and only if
$z\in\mathcal Z_{\tau}(\omega)$.

If $\bar z_{\phi}(\omega)\notin\mathcal Z_{\tau}(\omega)$,
$\mathcal R$ adjusts only generator-related setpoints and enforces
reactive-power limits via PV-to-PQ switching through four actions:
(i) fix the reactive-power outputs of violating non-reference
generators at their corresponding bounds and convert the associated
buses from PV to PQ;
(ii) redistribute a reference-generator active-power violation according
to the available active-power headroom of the remaining generators;
(iii) correct voltage violations through nearby generator-voltage
setpoints using clipped online secant sensitivities; and
(iv) relieve branch overloads through active-power redispatch
guided by signed power transfer distribution factor (PTDF) sensitivities of the active topology. The topology, transformer taps, and shunt settings remain
fixed.

For each trial control $u'$, let
$z':=z_{64}^{\mathrm{pin}}(u';\omega)$ denote its FP64 completion with
reactive-power-limit enforcement via PV-to-PQ switching. Relative to the incumbent point $z$, the trial is
accepted if $z'\in\mathcal Z_{\tau}(\omega)$ or
$\mathcal V_{\tau}(z';\omega)
<\zeta\mathcal V_{\tau}(z;\omega)$, where
$0<\zeta<1$; otherwise, the control adjustment is repeatedly halved.
Restoration terminates upon feasibility, failure to accept a trial, or
exhaustion of the iteration budget. Unsuccessful cases are retained and
counted as infeasible.

\begin{table*}[t]
\centering
\caption{Normal-case AC-OPF performance across seven heterogeneous
systems. Performance metrics are in percent.
$N_{\mathrm{ineq}}
:=\sum_{k\in\{P_g,Q_g,V,S_f,S_t\}}|\mathcal C_k|$
denotes the number of evaluated scalar inequality quantities.}
\label{tab:basecase_summary}
\setlength{\tabcolsep}{1.7pt}
\renewcommand{\arraystretch}{1.08}
\footnotesize

\begin{tabular}{@{}lr ccc ccc ccccc ccccc@{}}
\toprule
\multirow{2}{*}{System}
& \multirow{2}{*}{$N_{\mathrm{ineq}}$}
& \multicolumn{3}{c}{DeepOPF-U}
& \multicolumn{3}{c}{HH-MPNN}
& \multicolumn{5}{c}{CANOS}
& \multicolumn{5}{c}{UNION} \\
\cmidrule(lr){3-5}
\cmidrule(lr){6-8}
\cmidrule(lr){9-13}
\cmidrule(lr){14-18}
&
& Gap & CSR & IFR
& Gap & CSR & IFR
& Gap & CSR & IFR & Gap$_{\mathrm R}$ & IFR$_{\mathrm R}$
& Gap & CSR & IFR & Gap$_{\mathrm R}$ & IFR$_{\mathrm R}$ \\
\midrule

IEEE 57
& 231
& 2.46 & 72.43 & 0.00
& 2.39 & 75.44 & 0.00
& 1.14 & 96.87 & 6.20 & 0.78 & 100.00
& 1.68 & 99.54 & 72.00 & 1.63 & 100.00 \\

IEEE 118
& 598
& 15.58 & 71.68 & 0.00
& 7.13 & 72.65 & 0.00
& 0.14 & 97.06 & 0.00 & 0.21 & 100.00
& 0.88 & 99.97 & 72.50 & 0.89 & 100.00 \\

PEGASE-1354
& 5{,}856
& 58.14 & 72.75 & 0.00
& 71.20 & 74.53 & 0.00
& 2.22 & 95.47 & 0.00 & 2.68 & 98.20
& 2.38 & 99.34 & 0.00 & 2.51 & 100.00 \\

GOC-2312
& 8{,}778
& 35.28 & 72.50 & 0.00
& 34.87 & 74.04 & 0.00
& 7.94 & 98.22 & 0.00 & 8.48 & 19.95
& 1.96 & 99.92 & 0.00 & 1.96 & 96.95 \\

GOC-3970
& 17{,}498
& 29.45 & 71.62 & 0.00
& 27.09 & 75.42 & 0.00
& 9.05 & 95.85 & 0.00 & 9.40 & 100.00
& 0.65 & 99.66 & 0.00 & 0.79 & 100.00 \\

GOC-4601
& 19{,}265
& 35.34 & 71.71 & 0.00
& 22.23 & 74.85 & 0.00
& 6.32 & 94.52 & 0.00 & 6.30 & 99.90
& 0.58 & 99.75 & 0.00 & 0.58 & 100.00 \\

Korea-4492
& 16{,}794
& 3.32 & 73.61 & 0.00
& 7.58 & 76.37 & 0.00
& 4.01 & 97.88 & 0.00 & 3.99 & 100.00
& 0.25 & 99.96 & 46.20 & 0.26 & 100.00 \\

\midrule
Average
& --
& 25.65 & 72.33 & 0.00
& 24.64 & 74.76 & 0.00
& 4.40 & 96.55 & 0.89 & 4.55 & 88.29
& \textbf{1.20} & \textbf{99.73} & \textbf{27.24}
& \textbf{1.23} & \textbf{99.56} \\
\bottomrule
\end{tabular}
\vspace{-2mm}
\end{table*}

% ===============================
% Experiments
% ===============================
\section{Experiments}
\label{sec:experiments}

\subsection{Experimental Setup}

We evaluate UNION on seven AC-OPF systems:
IEEE~57, IEEE~118, PEGASE-1354, GOC-2312, GOC-3970,
GOC-4601, and Korea-4492, a real-world transmission grid from
the Korea Power Exchange (KPX). The benchmark set comprises
two IEEE cases, four large-scale PGLib v23.07 systems,
and one real-world grid, selected to span diverse network scales
under a common load-perturbation protocol
\cite{babaeinejadsarookolaee2021powergrid}.

Instances are generated by perturbing active and reactive
demands within $\pm10\%$ of nominal values. For each system,
we use 1{,}000 training, 1{,}000 validation, and 2{,}000 test
instances. UNION is trained once in
the shared seven-system setting, with pre-restoration validation CSR
used for early stopping and checkpoint selection over at most
7{,}000 epochs.

The model uses a three-layer EA-GNN with 64-dimensional embeddings,
graph-filter order $K=16$, and dropout $0.1$. Adam is used in FP32
with a batch size of 16 and a learning rate of $10^{-3}$. After a
20-epoch warm-up, the dual variables are updated using
$\rho=0.005$, with the update interval initialized to 10 epochs and
increased by five epochs thereafter. ECC uses
$\eta_{\beta}=10^{-3}$, $\mu_{\mathrm{cons}}=0.01$,
$\varepsilon=10^{-3}$, consensus-gradient clipping of $0.1$, and
the initialization $\alpha=0.5$. The training-time FP32 implicit
layer uses at most five Newton iterations with a tolerance of
$10^{-2}$, and its sparse linear systems are solved by cuDSS.
The full training run takes 175 hours on a single NVIDIA GeForce
RTX~4090 GPU.

\textit{Evaluation Metrics:}
Let $\mathcal T$ denote an evaluation set. Unless otherwise stated, cost accuracy is measured by the absolute
relative objective gap
\begin{equation}
\mathrm{Gap}(\mathcal T)
=
\frac{100}{|\mathcal T|}
\sum_{\omega\in\mathcal T}
\left|
\frac{\hat C(\omega)-C^\star(\omega)}
{C^\star(\omega)}
\right|,
\label{eq:eval_gap}
\end{equation}
where $\hat C(\omega)$ and $C^\star(\omega)$ denote the predicted
and reference objective costs for instance $\omega$, respectively.

For feasibility evaluation, we consider the following constraint
categories
\begin{equation}
\mathcal K
=
\{
P_g,\,
Q_g,\,
V,\,
S_f,\,
S_t,\,
P_{\mathrm{bal}},\,
Q_{\mathrm{bal}}
\},
\end{equation}
corresponding to generator active- and reactive-power limits,
voltage-magnitude limits, apparent-flow limits at both ends of each
branch, and active- and reactive-power balance residuals.
For each scalar constraint $c$, let
$v_c(\hat z(\omega);\omega)\ge 0$ denote its violation magnitude,
defined as the positive-part bound residual for an inequality
constraint and the absolute residual for a power-balance constraint.
Unless otherwise stated, the common evaluation tolerance is
$\tau=10^{-4}$~p.u.

Let $\mathcal C_k(\omega)$ denote the set of evaluated scalar
constraints in category $k$ for instance $\omega$. The constraint
satisfaction ratio (CSR) is defined as
\begin{equation}
\mathrm{CSR}(\mathcal T)
=
\frac{100}{|\mathcal K|}
\sum_{k\in\mathcal K}
\frac{
\displaystyle
\sum_{\omega\in\mathcal T}
\sum_{c\in\mathcal C_k(\omega)}
\mathbf{1}
\!\left[
v_c(\hat z(\omega);\omega)\le\tau
\right]
}{
\displaystyle
\sum_{\omega\in\mathcal T}
|\mathcal C_k(\omega)|
}.
\label{eq:eval_csr}
\end{equation}
Thus, CSR is the unweighted mean of the seven category-wise
scalar constraint-satisfaction ratios.

The instance feasibility rate (IFR) is defined as
\begin{equation}
\mathrm{IFR}(\mathcal T)
=
\frac{100}{|\mathcal T|}
\sum_{\omega\in\mathcal T}
\mathbf{1}
\left[
\max_{k\in\mathcal K}
\max_{c\in\mathcal C_k(\omega)}
v_c(\hat z(\omega);\omega)
\le \tau
\right].
\label{eq:eval_ifr}
\end{equation}
Hence, IFR is a strict instance-level metric; an instance is counted
as feasible only when all evaluated constraints are simultaneously
satisfied.

\textit{Evaluation Stages:}
For completion-compatible methods, unsubscripted metrics are computed
at the FP64-refined pre-restoration output, whereas
$\mathrm{Gap}_{\mathrm R}$ and $\mathrm{IFR}_{\mathrm R}$ are computed
at the restored output (Section~\ref{subsec:restoration}).
Direct-prediction baselines are evaluated at their native outputs;
failed restorations are counted as infeasible.

\begin{table*}[h]
\centering
\caption{Zero-shot performance on selected $N\!-\!1$ outages.
CSR is pre-restoration. Gap$_{\mathrm R}$ and IFR$_{\mathrm R}$
are post-restoration (\%).}
\label{tab:n1_summary}
\setlength{\tabcolsep}{3pt}
\renewcommand{\arraystretch}{1.08}
\footnotesize
\resizebox{\textwidth}{!}{%
\begin{tabular}{ll|ccc|cc|cc|ccc|ccc}
\toprule
\multirow{3}{*}{Case} & \multirow{3}{*}{System}
& \multicolumn{3}{c|}{Per-system-trained baseline}
& \multicolumn{7}{c|}{Jointly trained baselines}
& \multicolumn{3}{c}{Proposed unified model} \\
&
& \multicolumn{3}{c|}{PG-GNN \cite{yang2024pggnnopf}}
& \multicolumn{2}{c|}{DeepOPF-U \cite{liang2023deepopfuunifieddeepneural}}
& \multicolumn{2}{c|}{HH-MPNN \cite{arowolo2025generalizationgraphneuralnetworks}}
& \multicolumn{3}{c|}{CANOS \cite{piloto2024canosfastscalableneural}}
& \multicolumn{3}{c}{UNION} \\
& & Gap$_{\mathrm{R}}$ & CSR & IFR$_{\mathrm{R}}$ & Gap & CSR & Gap & CSR & Gap$_{\mathrm{R}}$ & CSR & IFR$_{\mathrm{R}}$ & Gap$_{\mathrm{R}}$ & CSR & IFR$_{\mathrm{R}}$ \\
\midrule
\multirow{4}{*}{Line outages} & IEEE 57 & 2.14 & 98.04 & 71.71 & 2.23 & 72.47 & 2.74 & 75.48 & \textbf{0.83} & 95.53 & 71.71 & 1.74 & \textbf{98.79} & \textbf{74.00} \\
 & IEEE 118 & 1.26 & \textbf{99.62} & 94.95 & 15.54 & 71.68 & 6.73 & 72.65 & \textbf{0.71} & 96.62 & 94.75 & 1.26 & 99.55 & \textbf{95.00} \\
 & GOC-4601 & 1.45 & 99.66 & \textbf{100.00} & 35.19 & 71.71 & 22.20 & 74.85 & 6.29 & 94.54 & \textbf{100.00} & \textbf{0.58} & \textbf{99.75} & \textbf{100.00} \\
 & Korea-4492 & 2.43 & 99.77 & \textbf{100.00} & 3.14 & 73.62 & 7.79 & 76.36 & 4.00 & 97.89 & \textbf{100.00} & \textbf{0.24} & \textbf{99.98} & \textbf{100.00} \\
\midrule
\multirow{4}{*}{Generator outages} & IEEE 57 & 1.50 & 97.29 & 85.75 & 3.62 & 72.43 & 3.21 & 75.44 & \textbf{0.43} & 95.06 & 85.75 & 1.13 & \textbf{98.20} & \textbf{100.00} \\
 & IEEE 118 & 1.24 & 98.05 & \textbf{70.07} & 12.03 & 71.69 & 6.83 & 72.66 & \textbf{1.02} & 96.26 & \textbf{70.07} & 1.45 & \textbf{98.32} & \textbf{70.07} \\
 & GOC-4601 & 1.75 & 99.59 & 90.00 & 33.28 & 71.72 & 20.33 & 74.85 & 6.50 & 94.49 & 99.90 & \textbf{0.93} & \textbf{99.63} & \textbf{100.00} \\
 & Korea-4492 & 2.40 & 99.81 & \textbf{100.00} & 3.50 & 73.62 & 7.41 & 76.35 & 3.84 & 97.77 & \textbf{100.00} & \textbf{0.27} & \textbf{99.86} & \textbf{100.00} \\
\bottomrule
\end{tabular}}
\vspace{-2mm}
\end{table*}

\subsection{Normal-Case Performance Across Heterogeneous Systems}
\label{subsec:normal_case}

Table~\ref{tab:basecase_summary} reports normal-case performance
on 14{,}000 held-out instances (2{,}000 per system). Each method
uses one model trained across all seven systems, \textit{without} per-system
retraining or test-time adaptation. It also reports
$N_{\mathrm{ineq}}$, the number of evaluated scalar inequality
quantities. For UNION, FP64 completion reduces active- and
reactive-power-balance residuals to $10^{-12}$--$10^{-9}$~p.u.,
yielding 100\% satisfaction for both equality categories at
$\tau=10^{-4}$~p.u. Thus, $N_{\mathrm{ineq}}$ counts
the inequality quantities that remain nontrivial after physical
completion, whereas
CSR and IFR are evaluated over all seven constraint categories in
Section~IV-A.

Before restoration, UNION achieves the lowest average Gap (1.20\%)
and the highest average CSR (99.73\%). However, IFR is low (27.24\%) because it is measured before restoration. UNION also
has the lowest Gap on the four largest systems: GOC-2312, GOC-3970,
GOC-4601, and Korea-4492. After restoration, the average
Gap$_{\mathrm R}$ and IFR$_{\mathrm R}$ are 1.23\% and 99.56\%,
respectively, with 100\% IFR$_{\mathrm R}$ on six systems and
96.95\% on GOC-2312, which implies restoration does not sacrifice cost. Under the same restoration operator,
CANOS attains an average IFR$_{\mathrm R}$ of 88.29\%.

The largest distinction is observed on GOC-2312. Under the same
restoration operator, UNION and CANOS attain IFR$_{\mathrm R}$ values
of 96.95\% and 19.95\%, respectively. CANOS's 98.22\%
pre-restoration CSR, despite its low IFR$_{\mathrm R}$, shows that
scalar-level satisfaction does not capture the severity, structure, or
reachability of the remaining violations under the available generator
controls. An analysis of the GOC-2312 case data shows that 20.9\% of
its rated branches have limits at or below 50~MVA, whereas achievable
flow corrections depend on generator headroom and signed PTDF
sensitivities. A post-hoc analysis of UNION's restored test outputs
shows that its residual violations are confined to apparent-power limits
on 19 of 3{,}013 rated branches, with three accounting for most
unsuccessful instances. These include rated near-zero-impedance
couplers and high-impedance radial branches with limited redispatch
leverage. The remaining failures are therefore concentrated in a small
set of branch limits that are difficult to correct through
generator-only adjustments.

\begin{figure*}[htbp]
    \centering
    \subfloat[]{
        \includegraphics[width=0.48\textwidth]{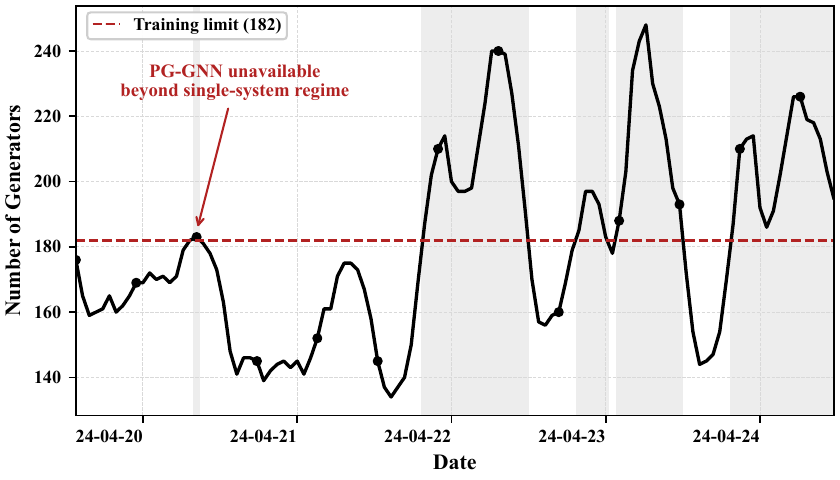}
        \label{fig:four_subplots_a}
    }
    \hfill
    \subfloat[]{
        \includegraphics[width=0.48\textwidth]{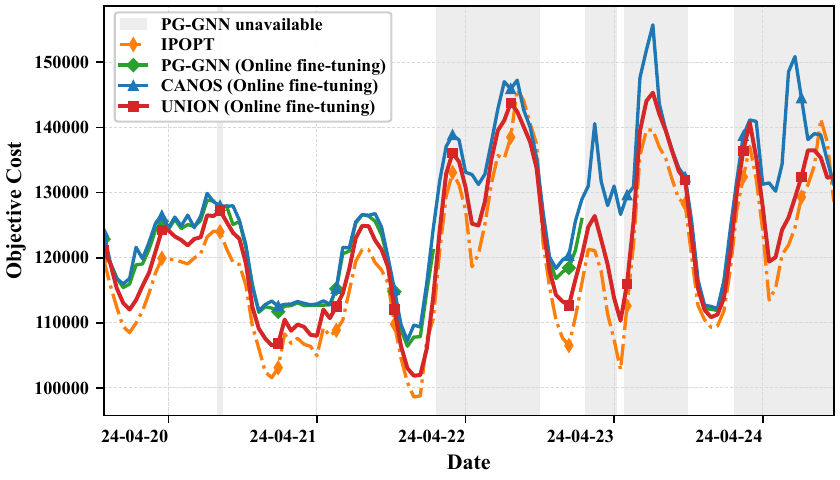}
        \label{fig:four_subplots_b}
    }
    \caption{Temporal validation on Korea-4492.
(a) active-generator count and single-system training limit.
(b) objective cost.}
    \label{fig:four_subplots}
\end{figure*}

\subsection{Zero-Shot Performance Under $N\!-\!1$ Contingencies}
\label{subsec:n1}

Table~\ref{tab:n1_summary} reports zero-shot performance on
selected $N\!-\!1$ contingencies for IEEE 57, IEEE 118,
GOC-4601, and Korea-4492. For each system, a stratified subset of 20 line outages and
up to 10 eligible non-reference generator outages is evaluated
using 1{,}000 load instances per contingency. Line outages are selected by ranking branches by pre-contingency
loading and selecting uniformly across the ranking.
Each generator outage fully removes one eligible non-reference
modeled generator, including its active- and reactive-power
capabilities and voltage-regulation action; two reference-infeasible
IEEE 57 outages are excluded. All retained
cases are evaluated without contingency-specific retraining or
adaptation. These experiments therefore assess zero-shot
generalization over the evaluated contingency set rather than
constituting an exhaustive $N\!-\!1$ security assessment.
PG-GNN is trained per system, whereas UNION and the jointly
trained baselines use one model across all seven systems; the
native-output IFRs of DeepOPF-U and HH-MPNN are below
0.1\% and are excluded.

UNION attains 100\% IFR$_{\mathrm R}$ on GOC-4601 and
Korea-4492 under both outage types, with Gap$_{\mathrm R}$ values
of 0.24\%--0.93\%. This performance is consistent with their
corrective redundancy; 133 and 182 generators provide active- and reactive-power control while their meshed
networks offer multiple paths for PTDF-guided redispatch.
Consequently, even large outage-induced imbalances can be distributed
across many controls rather than concentrated at a few units or
branches.

\begin{table}[h]
\centering
\caption{Temporal validation on 116 hourly Korea-4492 snapshots. All values are in percent.}
\label{tab:korea4492_temporal}
\setlength{\tabcolsep}{2.5pt}
\renewcommand{\arraystretch}{1.06}
\footnotesize
\begin{tabular}{@{}llcccc@{}}
\toprule
\multirow{2}{*}{Method}
& \multirow{2}{*}{Setting}
& \multirow{2}{*}{Coverage}
& \multicolumn{1}{c}{Pre-rest.}
& \multicolumn{2}{c}{Restored output} \\
\cmidrule(lr){4-4}
\cmidrule(lr){5-6}
&
& & CSR & Gap$_{\mathrm R}$ & IFR$_{\mathrm R}$ \\
\midrule

\multirow{2}{*}{PG-GNN~\cite{yang2024pggnnopf}}
& Zero-shot
& 56.03 & 96.19 & 3.55 & 70.8 \\
& Online fine-tuning
& 56.90 & 97.09 & 3.46 & 71.2 \\

\addlinespace[1.5pt]

\multirow{2}{*}{CANOS~\cite{
piloto2024canosfastscalableneural}}
& Zero-shot
& 86.21 & 95.72 & 4.02 & 64.0 \\
& Online fine-tuning
& \textbf{100.00} & 97.72 & 4.22 & 75.0 \\

\addlinespace[1.5pt]

\multirow{2}{*}{UNION}
& Zero-shot
& \textbf{92.24} & 96.40 & 2.68 & 72.9 \\
& Online fine-tuning
& \textbf{100.00}
& \textbf{98.73}
& \textbf{2.51}
& \textbf{82.8} \\

\bottomrule
\end{tabular}
\vspace{-2mm}
\end{table}

\begin{table*}[t]
\centering
\caption{Pre-restoration ablation of graph aggregation and ECC;
Gap and CSR are averaged equally across seven systems.}
\label{tab:ablation_consensus}
\setlength{\tabcolsep}{4pt}
\renewcommand{\arraystretch}{1.08}
\footnotesize
\begin{tabular}{lcc|cc|cc|cc|cc}
\toprule
\multirow{2}{*}{Aggregation}
& \multicolumn{2}{c|}{Global Mean \eqref{eq:mean_pool}}
& \multicolumn{2}{c|}{Local Attention \eqref{eq:attn_pool}}
& \multicolumn{2}{c|}{Static Mixture $(\beta=0)$}
& \multicolumn{2}{c|}{SGA without ECC}
& \multicolumn{2}{c}{SGA + ECC} \\
& Gap (\%) & CSR (\%)
& Gap (\%) & CSR (\%)
& Gap (\%) & CSR (\%)
& Gap (\%) & CSR (\%)
& Gap (\%) & CSR (\%) \\
\midrule
Overall
& 1.95 & 99.32
& 1.78 & 99.00
& 1.66 & 99.58
& 1.43 & 99.65
& \textbf{1.19} & \textbf{99.74} \\
\bottomrule
\end{tabular}
\vspace{-2mm}
\end{table*}

The lower IFR$_{\mathrm R}$ values on the IEEE systems arise from
different limiting constraints. Under IEEE 57 line outages, the
remaining failures are voltage-limited; with only seven
generator-voltage setpoints, restoration cannot always eliminate the
shallow residuals (median 0.0023~p.u.). Under IEEE 118 generator
outages, unresolved cases are dominated by deeper apparent-flow
violations, with median residuals of 0.185 and 0.138~p.u. at the two
branch ends. A generator loss requires substantial transfer redistribution, but the
available network paths and PTDF leverage are insufficient to eliminate
all resulting branch overloads. The identical 70.07\%
IFR$_{\mathrm R}$ obtained by PG-GNN, CANOS, and UNION in this
case further indicates a system-level restoration limit rather than
a predictor-specific failure. Overall, restoration success is
governed by violation severity relative to the available corrective
degrees of freedom, not by network size alone.

\subsection{Real-World Temporal Adaptation on Korea-4492}

We evaluate 116 hourly Korea-4492 snapshots collected over five days with time-varying generator availability and active line states. Each
method is evaluated zero-shot and with online fine-tuning within the
same 5-min operating interval. At each update, 20 AC-OPF samples are
generated in approximately 40~s from the operating condition available
5~min before the target time; neural fine-tuning is limited to 1~min.
The fine-tuning set accumulates over time while evaluation uses the
actual target-time snapshot, whose active topology may differ from
that used for sample generation.

Fig.~\ref{fig:four_subplots} and
Table~\ref{tab:korea4492_temporal} summarize the results. Coverage
denotes the fraction of snapshots for which the complete inference
pipeline produces a power-flow-consistent operating point. PG-GNN
becomes unavailable when active generators exceed its
fixed 182-generator output configuration, as shown in
Fig.~\ref{fig:four_subplots}(a). CANOS and UNION remain dimensionally
compatible across all snapshots, although zero-shot FP64 power-flow
refinement may fail for some predicted controls. Because
Gap$_{\mathrm R}$, CSR, and IFR$_{\mathrm R}$ are computed only on
covered snapshots, they must be interpreted jointly with coverage.
Objective gaps are computed relative to IPOPT
solutions.

Zero-shot UNION provides the highest coverage at 92.24\%, compared
with 86.21\% for CANOS and 56.03\% for PG-GNN. Online fine-tuning raises CANOS and UNION to 100\%, whereas PG-GNN
remains limited to 56.90\% by its fixed output dimension. Fine-tuned UNION attains a
Gap$_{\mathrm R}$ of 2.51\% and an IFR$_{\mathrm R}$ of 82.8\%,
with a 1.71-percentage-point lower Gap$_{\mathrm R}$ and a
7.8-percentage-point higher IFR$_{\mathrm R}$ than CANOS.

Uncovered zero-shot UNION snapshots occur when the predicted
active-power dispatch cannot be balanced within the reference-generator
range, preventing the power-flow solve from converging and blocking
restoration. Full coverage after brief fine-tuning therefore suggests
that operating-level recalibration is central to recovery despite
continued topology variation. On covered snapshots, the common
restoration operator largely removes generator active- and
reactive-power and branch-flow violations, while residual infeasibility
is concentrated in voltage-limit violations at weakly controllable
radial buses behind fixed off-nominal transformer taps.
Fig.~\ref{fig:four_subplots}(b) shows that online fine-tuning preserves
the temporal cost trend, while
Table~\ref{tab:korea4492_temporal} reports improved coverage and strict
feasibility.

\subsection{Ablation Study of UNION}
\label{subsec:ablation}

Table~\ref{tab:ablation_consensus} compares the graph-level
aggregation configurations for UNION. All variants share the same
encoder, control predictor, implicit layer, and primal--dual objective,
and differ only in the aggregation rule and use of ECC. Each is
independently trained, selected using pre-restoration CSR on the common
validation split in Section~IV-A, and evaluated using held-out test Gap
and CSR averaged equally across the seven systems. Restoration is
excluded so that deterministic post-processing does not obscure
differences in the learned predictor and training mechanism.

The two single-aggregation variants exhibit complementary behavior;
attention pooling reduces Gap relative to mean pooling
(1.78\% versus 1.95\%), whereas mean pooling yields higher CSR
(99.32\% versus 99.00\%). Their static mixture improves both metrics,
indicating that the uniform and attention summaries provide
complementary information. Learning the mixture through SGA further
reduces Gap to 1.43\% and increases CSR to 99.65\%. Adding ECC yields
the best pre-restoration results (1.19\% Gap and 99.74\% CSR). These
incremental improvements support SGA with ECC as the final configuration
and show the separate empirical contributions of uniform--attention
mixing, learned scalar gating, and gate-level consensus correction.

\begin{table}[h]
\centering
\caption{Mean single-instance latency at batch size one on the three
largest systems (ms).}
\label{tab:latency}
\setlength{\tabcolsep}{4pt}
\renewcommand{\arraystretch}{1.08}
\footnotesize
\resizebox{\columnwidth}{!}{%
\begin{tabular}{lccc}
\toprule
Method / stage & GOC-3970 & GOC-4601 & Korea-4492 \\
\midrule
MATPOWER (MIPS)              & 5,341  & 5,490  & 3,508 \\
PowerModels.jl (IPOPT)       & 10,827 & 11,467 & 5,630 \\
ExaModels+MadNLP+CPU (MA57)  & 3,543  & 3,439  & 1,356 \\
ExaModels+MadNLP+GPU (cuDSS) & 1,023  & 1,194  & 591 \\
\midrule
Dense implicit layer         & 225    & 258    & 259 \\
Sparse-aware implicit layer  & \textbf{55}     & \textbf{57}     & \textbf{58} \\
\midrule
Full UNION pipeline          & \textbf{114}    & \textbf{108}    & \textbf{108} \\
\bottomrule
\end{tabular}%
}
\vspace{-2mm}
\end{table}

\subsection{Computational Efficiency}

Table~\ref{tab:latency} reports mean per-instance latency at
batch size one on the three largest systems. MATPOWER (MIPS),
PowerModels.jl (IPOPT), and ExaModels+MadNLP+CPU (MA57) are
executed on CPU, whereas ExaModels+MadNLP+GPU (cuDSS) and all
UNION implementations are executed on GPU. Exploiting Jacobian
sparsity reduces the implicit-layer latency from 225--259~ms to
55--58~ms, a 4.1--4.5$\times$ speedup.

Including FP64 power-flow refinement and restoration, the full
UNION pipeline requires 108--114~ms, remaining below 0.12~s on
all three large-scale systems. Under this execution setting, it is
52--106$\times$ faster than PowerModels.jl (IPOPT) and
5.5--11.1$\times$ faster than ExaModels+MadNLP+GPU (cuDSS).

% ===============================
% Conclusion
% ===============================
\section{Conclusion}
\label{sec:conclusion}

This paper presented UNION, a unified graph--implicit AC-OPF
framework for heterogeneous systems and topology-varying operation.
Its key principle is a division of labor between learning and physics;
the shared graph model learns transferable generator-control setpoints,
while the sparse implicit layer recomputes the dependent AC state from
the nonlinear power-flow equations of the active topology. Thus, when
an outage changes the admittance matrix and Jacobian structure, UNION
does not extrapolate the full post-contingency state from the GNN
alone; it physically completes the state for the modified network.
This topology-conditioned completion underlies its zero-shot
$N\!-\!1$ performance. Without contingency-specific adaptation,
UNION achieved 100\% IFR$_{\mathrm R}$ on GOC-4601 and Korea-4492
under both line and generator outages, with
Gap$_{\mathrm R}$ values of 0.24\%--0.93\%. These results show that
shared multi-system learning can be combined with active-topology
physics to support reliable real-time AC-OPF under unseen operating
configurations.

% ===============================
% Acknowledgment
% ===============================
\section*{Acknowledgment}
This work was supported by the National Research Foundation of Korea (NRF) under Grant RS-2025-02215243 and the Korea Institute of Energy Technology
Evaluation and Planning (KETEP) under Grant RS-2026-25527712.
% ===============================
% References
% ===============================
\bibliographystyle{IEEEtran}
\bibliography{ref}

\end{document}